 \newlength\smallfigwidth
 \documentclass[aps,prb,reprint,floatfix,showpacs,amsmath,amssymb]{revtex4-1}

\usepackage{graphicx}
\def\ba{\begin{eqnarray}}
\def\ea{\end{eqnarray}}
\def\be{\begin{equation}}
\def\ee{\end{equation}}

\begin{document}

\preprint{UFV}

\title{A note on topological invariants in condensed matter}

\author{J. M. Fonseca}
\email{jakson.fonseca@ufv.br}
 \affiliation{ Universidade Federal de Vi\c cosa, Departamento de F\'isica, 
Avenida Peter Henry Rolfs s/n, 36570-000, Vi\c cosa, MG, Brasil\\
}

\author{V. L. Carvalho-Santos}
\email{vagson.santos@ufv.br}
\affiliation{Instituto Federal de Educa\c c\~ao, Ci\^encia e Tecnologia Baiano - Campus 
Senhor do Bonfim, \\Km 04 Estrada da Igara, 48970-000 Senhor do Bonfim, Bahia, Brazil}
\affiliation{Departamento de F\'isica, Universidad de Santiago de Chile and CEDENNA,\\ Avda. Ecuador 3493, Santiago, Chile}

\date{\today}

\begin{abstract}

We discuss some aspects of topological invariants that classify topological states of matter with emphasis on topological insulators. The main aspect addressed is if there are only two topological phases to Bloch Hamiltonian that are time reversal invariant or if there are more phases that have different topological invariants. From a mathematical point of view may exist more topological phases of matter
as a subclass of one well established phase.

\vspace{.5cm}
\noindent
{\bf keywords}: topological insulators, topological inavariants, topology, topological states of the matter

\end{abstract}



%

\maketitle


The classification of distinctive phases of matter is an important and recurring theme in condensed matter physics. For several condensed matter systems (CMS), the Landau theory of phase transitions, in which the states are characterized by underlying symmetries that are spontaneously broken, is used to describe phase transitions. On the other hand, topological quantum states of the matter are exciting states that do not break any symmetry and can not be described by the Landau approach. Instead they are associated to the notion of topological order, described by topological quantum numbers which are, many times, associated with the bulk wavefunction that describes the system. As examples, we can cite the quantum Hall effect, \cite{Thouless-1982} topological insulators, \cite{Review1,TI review} topological superconductors and superfluids. \cite{shen} These CMS do not break any symmetry, but they define a topological phase since some fundamental properties are insensitive to smooth changes in material parameters and can not change unless the system pass through a quantum phase transition. In this context, there is a considerable interest in understanding the topological quantum states of the matter due their potential for producing new physical phenomena as well as future technological applications. 

Regarding topological insulators, a lot of effort have been done, as experimentally as theoretically in order to understand their properties. From the theoretical point of view, different mathematical formulations have been developed to obtain the $Z_2$ topological insulator, $\nu$.
\cite{Kane-PRL-2005, fu2006, fu2007, Bernevig-PRL-2006,fukui2007,moore,fukui2008,Qi, roy, wang} Some of them are more useful to computational calculus, others to physical interpretations, but all they are mathematically and physically equivalents. These theoretical models predict a phase transition from a trivial insulator to the quantum spin Hall insulator. In order to cover the normal and the inverted band structure, HgTe quantum wells were grown. \cite{Konig-Science-2007} It has been shown that when the quantum well attains a critical thickness $d_{QW}>d_c$, the band structure is inverted, indicating a negative energy gap predicted in the model of Bernevig \textit{et al}. \cite{Bernevig-PRL-2006} This nontrivial inversion band structure was also detected optically. \cite{Schmidt-PRB-2009} In addition, by using angle-resolved photoemission spectroscopy, surface states with a single Dirac cone, which characterizes a three dimensional topological insulator, have been observed in a class of materials. \cite{Xia-2009,Zhang-2009,Chen-2009,Hiesh-2009}

However, despite of this recent research field find a rapid experimental and theoretical success, it is important to stablish some aspects in a solid mathematical bases in order to understand the topological equivalence among these states from the viewpoint of homotopy theory. In this context, the purpose of this article is to discuss some aspects of topological invariants that classify topological states of matter, in particular, topological insulators.

Topological invariants are quantities which are conserved under homeomorphisms. \cite{nakahara} Homeomorphism is a mapping denoted by $f\,:\, X_1\rightarrow X_2$ which is continuous and has an inverse mapping $f^{-1}\,:\, X_2\rightarrow X_1$ also continuous. When exist a homeomorphism between topological spaces $X_1$ and $X_2$ we say that $X_1$ is homeomorphic to $X_2$. One can show that a homeomorphism is an equivalence relation $\sim$  satisfying the properties \cite{nakahara}

\begin{description}
\item[(i)] a $\sim $ a,  reflective;
\item[(ii)] If a $\sim$ b, then b $\sim$ a, symmetric; 
\item[(iii)] If a $\sim$ b and b $\sim$ c, then a $\sim$ c, transitive;
\end{description}

\noindent
then, one can divide all topological spaces into equivalence classes according to whether it is possible to 
continuous map one space into the other by a homeomorphism. Intuitively, two topological spaces are homeomorphic 
each other if we can continuously map (like we can ``deform rubber'') into the other (without tearing). 
Condensed matter physics classifies all time reversal invariant insulators in two topological classes.
Considerer the Bloch Hamiltonian that describes free electrons in a periodic potential
produced by ions in a cristaline lattice. This Hamiltonian is given by sum of kinetic and
potential energy, where the potencial energy has the lattice symmetry 
$U(\mathbf r + \mathbf R) =U(\mathbf r)$, where $\mathbf R$ is a Bravais lattice vector. 
The eigenstates $\psi$ can be chosen to have the form
of a plane wave times a function with the periodicity of the Bravais lattice:
\be
\psi_{n\mathbf k}(\mathbf r)=e^{i\mathbf k \cdot \mathbf r} u_{n\mathbf k}(\mathbf r)\,,
\ee
where $u_{n\mathbf k}(\mathbf r + \mathbf R) =u_{n\mathbf k}(\mathbf r)$. The Bloch wavefunctions
$u_{n\mathbf k}(\mathbf r)$ for the ocuppied states in the bulk of cristal determines
the topological properties of the material. To be more specific, we, can look to the properties
of the bloch functions in high symetry points in the Brilloun zone, like the parity
of the $| u_{n\mathbf k}(\mathbf k) \rangle $ (when the cristal has parity inavariance 
in addition time reversal invariance) and this properties given the topological
class of the material. This topological class can be of two types, an ordinary or
trivial insulator and a topological insulator. In the surface of a topological insulator
there are electronic states without an energy gap, because the bulk has a energy gap,
the topological invariant that classifies the topological insulator phase has to
change in the surface, else the topological phase does not change between a material
and the vacuum (a trivial insulator), then the presence of electronic states
without an energy gap in the surface is a consequence of bulk properties. This is
a bulk-boundary correspondence like in the quantum Hall effect. \cite{TI review} 

Consider a ${\cal T}$ invariant Bloch Hamiltonian,
which must satisfy

\be
\Theta \, {\cal H}(\mathbf k) \, \Theta^{-1}={\cal H}(-\mathbf k)\,.
\ee

\noindent
We can considere the equivalence classes of Hamiltonians satisfying this constraint
imposed by TRS that can be smoothly deformed without closing the energy gap. In others
words, the class of Hamiltonians that are homeomorphic and can be mapped each other.
Mathematically there are many topological quantities that can be used to classify 
these equivalence classes. Physics use a topological quantity $\nu$ that can assume two
possible values, $0$ (even) to trivial or ordinary insulator and $1$ (odd) 
to topological insulators. \cite{TI review} 
There are physical arguments
to support the existence of only two topological classes, so called $Z_2$ topological classification. 
These arguments apply to two dimensional insulators with an energy gap between the valence and conduction 
energy bands and they can be generalised for three dimensional insulators with an energy gap. \cite{TI 
review}

It should be emphasized that the physicists use only {\bf one} topological invariant 
to classify the topological classes of the matter. On the other hand the 
mathematicians do not know how we can characterize completely the equivalence class of homeomorphism \cite{nakahara} like the equivalence class of time reversal Hamiltonians that are time reversal symmetrical, i.e. find all topological invariants that homeomorphism.
From the mathematical point of view, there is only a partial answer to this question, in such way that what we can say is that if two spaces have different topological invariants they are not homeomorphic to each other. However, we do not know how to specify all topological invariants in a homeomorphism, nevertheless we know only a partial set of topological invariants. 

The topological classification used currently by condensed matter physicists is not affected by the discussion presented here. However, it is important to the physicists understand that
at in principle, one topological state or phase of the matter that is characterized by one
topological invariant (like one strong topological insulator with $\nu=1$) can
contain many distinct topological states of the matter and maybe can present different
physical effects associated to this distinct topological states. Therefore, one can classify
one topological phase in others topological subclasses.

One example can be providing by a three dimensional topological insulator. \cite{TI review} In analogy to a two dimensional topological insulator we have one $Z_2$ topological number $\nu_0$ that specifies if the topological insulator is a strong or a weak one, but we have more three topological numbers $(\nu_1,\,\nu_2,\,\nu_3)$ that specify the subclasses. For example, if a weak topological insulator, $\nu_0=0$, has the numbers $(\nu_1,\,\nu_2,\,\nu_3)=(1,\,1,\,1)$, it have some topological properties presented by a strong topological insulator, but this state can be destroyed for small perturbations. \cite{TI review}

In conclusion we can say: {\it if two topological spaces (Bloch Hamiltonian) have different
topological invariants they can not be homeomorphic to each other and if
two spaces have the same topological invariant (like trivial insulator $\nu =0$,  or 
topological insulator $\nu=1$). They can be in different equivalence class and can be not
stay in the same topological class}. Are there only two topological classes to
time reversal invariant Block Hamiltonians with a gap? if there are more than two
classes are this suclassificabtion important to condensed matter physical point of view?
A definitive answer to this question will be important to physical interpretation and
discussion of topological states of the matter and a profound and enlightening from a mathematical point of view. 

{\it Note}: The authors do not know any similar discussion to this presented here in the 
physical literature.

\newpage
\centerline{\bf Acknowledgments}
The authors thank D.H.T. Franco for useful discussions
They are also grateful to FAPEMIG, Capes and CNPq (Brazilian agencies) for financial support.

\end{document}